\def \SAIT #1 #2 {{\em Mem.\ Soc.\ Astron.\ It.\/} {\bf #1}, #2}
\def \MESS #1 #2 {{\em The Messenger\/} {\bf #1}, #2}
\def \ASTRNACH #1 #2 {{\em Astron. Nach.\/} {\bf #1}, #2}
\def \AAP #1 #2 {{\em Astron. Astrophys.\/} {\bf #1}, #2}
\def \AAL #1 #2 {{\em Astron. Astrophys. Lett.\/} {\bf #1}, L#2}
\def \AAR #1 #2 {{\em Astron. Astrophys. Rev.\/} {\bf #1}, #2}
\def \AAS #1 #2 {{\em Astron. Astrophys. Suppl. Ser.\/} {\bf #1}, #2}
\def \AJ #1 #2 {{\em Astron. J.\/} {\bf #1}, #2}
\def \ANNREV #1 #2 {{\em Ann. Rev. Astron. Astrophys.\/} {\bf #1}, #2}
\def \APJ #1 #2 {{\em Astrophys. J.\/} {\bf #1}, #2}
\def \APJL #1 #2 {{\em Astrophys. J. Lett.\/} {\bf #1}, L#2}
\def \APJS #1 #2 {{\em Astrophys. J. Suppl.\/} {\bf #1}, #2}
\def \APSS #1 #2 {{\em Astrophys. Space Sci.\/} {\bf #1}, #2}
\def \ASR #1 #2 {{\em Adv. Space Res.\/} {\bf #1}, #2}
\def \BAIC #1 #2 {{\em Bull. Astron. Inst. Czechosl.\/} {\bf #1}, #2}
\def \JSQRT #1 #2 {{\em J. Quant. Spectrosc. Radiat. Transfer\/} {\bf #1}, #2}
\def \MN #1 #2 {{\em Mon. Not. R. Astr. Soc.\/} {\bf #1}, #2}
\def \MEM #1 #2 {{\em Mem. R. Astr. Soc.\/} {\bf #1}, #2}
\def \PLR #1 #2 {{\em Phys. Lett. Rev.\/} {\bf #1}, #2}
\def \PASJ #1 #2 {{\em Publ. Astron. Soc. Japan\/} {\bf #1}, #2}
\def \PASP #1 #2 {{\em Publ. Astr. Soc. Pacific\/} {\bf #1}, #2}
\def \NAT #1 #2 {{\em Nature\/} {\bf #1}, #2}
\newcommand{\ltsima} {$\; \buildrel < \over \sim \;$}
\newcommand{\gtsima} {$\; \buildrel > \over \sim \;$}
\newcommand{\lta} {\lower.5ex\hbox{\ltsima}}
\newcommand{\gta} {\lower.5ex\hbox{\gtsima}}

\documentstyle[twoside]{memsait}
\input epsf.sty
\begin{opening}
\title{MODELS FOR THE X-RAY EMISSION FROM RADIO QUIET AGNs}
\author{Francesco Haardt}
\institute{Dipartimento di Fisica dell'Universit\'a degli Studi di Milano}
\date{}
\end{opening}

\begin{document}

%\oddpagefooter{\sf Mem. S.A.It., Vol. ??, ??}{}{\thepage}
%\evenpagefooter{\thepage}{}{\sf Mem. S.A.It., Vol. ??, ??}
\oddpagefooter{}{}{} % LEAVE AS IT IS !
\evenpagefooter{}{}{} % LEAVE AS IT IS !
\
\bigskip

\begin{abstract}
The current status of understanding of the X-ray emission from Seyfert galaxies
involves thermal Comptonization of soft photons by mildrelativistic electrons
and positrons. I review observational and theoretical arguments supporting such
a view, discussing current models proposed for the structure of the innermost
part of the accretion flow: extended coronae, small scale flaring blobs,
advection dominated accretion disks and small scale cold cloudlets. 

\end{abstract}

\section{Introduction}

One of the most remarkable aspect of AGNs is the strong X-ray emission, the
origin of which may be directly related to the central engine powering the
active galaxy. In fact the rapid X-ray variability detected in several sources
(e.g. McHardy 1989); indicates that the X-ray production region is very close
to the central object. The study of the mechanisms leading to such a strong
X-ray emission is then extremely important in view of understanding the
physical condition of the inner regions of AGNs. While in radio loud objects
the bulk of the emission seems to be non thermal (e.g. Ghisellini \& Madau
1996), in Seyfert galaxies there is evidence of the presence of cold thermal
matter near the central engine (Pounds et al. 1990). The accreting supermassive
black hole picture is widely accepted for unbeamed objects, and the strong UV
excess is thought to be the signature of the accretion flow. In this picture
the origin of the X-rays remains unclear, and any successful model needs to
satisfy three main observational constrains: 

\begin{enumerate}
\item{} the size of the X-ray production region must be small, few
gravitational radii, as indicated by the fast variability (McHardy 1989); 
\item{} the observed spectrum in the hard X-rays ([2-20] keV) is close to a
power law, with a small dispersion in the value of the spectral index, with
mean value of $\simeq 0.7 \pm 0.15$ for Seyfert galaxies (e.g. Mushotzky 1984).
At higher energies the power law shows a break. Force fitting an exponential
cut--off to the break, the e--folding energy is comprised between 50 and 300
keV (Madejski et al. 1995); 
\item{} whatever produces the X-rays is close to cold reflecting matter, as
inferred from the presence of fluorescent Fe line emission and Compton
reflection hump peaked around 30 keV. After deconvolution of the latter
component the spectral index of the underlying power law is revised to $\simeq
0.9$ (Pounds et al. 1990). 
\end{enumerate}

Two classes of models have been proposed to explain the shape of the X-ray
spectrum. One is based on the production of very high energy primaries and
strong reprocessing via electromagnetic cascades leading to the formation of an
$e^{\pm}$ pair plasma (e.g. Zdziarski et al. 1993). The second involves
multiple Compton scattering (Comptonization) of soft photons on a thermal
population of hot electrons (see Svensson 1996 for a recent review). 

In this contribution I shall discuss some aspects of the origin of the X--rays
in Seyfert galaxies. I will concentrate mainly on the radiation generation
mechanisms. Then I will review some current pictures of the innermost part of
the accretion flow in radio quiet AGNs. For a somewhat different approach the
reader may be interested in reading the review by Maraschi \& Haardt (1996). 

\section{Thermal or non Thermal X--rays?}

The X--rays emitted by AGNs have obviously a non--thermal spectrum. The
question addressed in the title above rather regards the particle distribution
responsible for such an emission, which is largely unknown and that can well be
thermal. Before speculating the nature of the particle distribution, it is
important to ask (and possibly answer to): how do particles cool down? 

Fast particles cool down by means of three possible interactions, namely
particle--particle, particle--photon and particle--magnetic field, transforming
kinetic energy into free--free (FF), inverse Compton (IC), and synchrotron (S)
radiation, respectively. Which one dominates remains to be seen. The concept of
"compactness" $\ell$ is useful to assess the importance of different radiation
mechanisms. The compactness is defined as 
\begin{equation}
\ell\equiv {\sigma_T \over m_ec^3}{L\over R} \simeq 10^{4} {{\cal{L}}\over r}
\end{equation}
where $\cal L$ is the luminosity in Eddington units, $r$ is the source size in
units of $2GM/c^2$, and the other symbols have their usual meaning. 

Neglecting for the moment S emission, and assuming that the gas is thermal, the
relative importance of FF vs. IC emission can be translated into a limit for
$\ell$. One can easily find that if 
\begin{equation}
\ell \gta 0.03/\sqrt{kT_e/m_ec^2} 
\end{equation}
IC cooling is much faster than FF cooling. Now the value of the compactness in
Seyfert galaxies is largely unknown, but rough estimates range from 1 to 300
(Done \& Fabian 1989). This, together with the OSSE results indicating
$kT_e\simeq 50-300$ keV, implies that IC cooling is largely dominant. 

Another nice property of the compactness parameter is immediately evident once
we notice that $\ell$ is proportional to the photon--photon opacity for pair
production. For $\ell \gta 60$ the hot gas cools down via radiation emission
{\it and} via mass ($e^+e^-$) production. From the observed values of $\ell$,
we argue that $e^+e^-$--pairs are an important ingredient of the X--ray
emitting gas in Seyfert galaxies. 

The role of S cooling, neglected so far, depends on the basically unknown value
of the magnetic field. Taking an equipartition field, the primary S emission
is, under the expected conditions, almost completely self--absorbed. In other
words, synchrotron photons may play a role as a soft input for Comptonization
(SSC models, see next), rather than being directly responsible for the cooling.

Now I will try to justify the assumption that the gas is thermal.

The main difference between thermal and non--thermal models consists in the
assumption made on the channelling of the available power. If the power is
equally channeled into all the electrons we talk about thermal plasma, while if
the power is channeled into a tiny fraction of the electrons we are dealing
with non--thermal plasma. Physically, one can think that in a thermal plasma
the cooling time is longer than the particle--particle collision time, the
contrary for a non--thermal plasmas. 

In the {\it thermal} Comptonization models (e.g. Sunyaev \& Titarchuk 1980),
the underlying assumption is that all the emitting particles are heated and
maintained in a Maxwellian distribution at constant temperature, despite the
rapid variations in luminosity that are observed. Density and temperature of
the electrons were usually assigned in a rather ad hoc way, in order to produce
a power law spectrum with the observed index, but this problem has been solved
considering the interplay between the cold accretion disk, and the hot corona
(Haardt \& Maraschi 1991, 1993; see also next \S 3). 

Non--thermal pair models (for a review see Svensson 1992, 1994) predict the
correct spectral index with the only assumption that $\ell \gta 60$, and then,
for this reason, they seemed to be favoured. On the other hand, after the OSSE
results was clear that the steepening of the hard X-ray spectrum above 100 keV
predicted by non--thermal models was only barely as sharp as the data required
(e.g. Maisack et al. 1993). Furthermore, the non--thermal pair model predicts a
flattening of the spectrum at $\sim 100-200$ keV, and the presence of a
conspicuous annihilation line, neither of which is observed. OSSE observations
instead can be interpreted in the framework of thermal Comptonization models.
If this is the case, also the X--ray background can be best explained as the
sum of the emission of Seyfert galaxies (see e.g. Madau et al. 1993, 1994). 

Finally I would like to recall that the observed rapid X--ray variability was
longer considered as an evidence {\it against} thermal models. However
Ghisellini et al. (1993) demonstrated that in the spectrum formation what
really matters is the average energy of the particles rather than the shape of
the particle distribution. No matter what the shape is, as long as the average
particle energy is small (as result of, e.g., pair cascade with particle
re--acceleration), the emitted IC spectrum is practically indistinguishable
from that emerging from a thermal particle distribution. 

However the particle distribution could be {\it genuinely thermal}. Ghisellini
et al. (1996) show that relativistic electrons (and positrons) can thermalize
in few synchrotron cooling times by emitting and absorbing cyclo--synchrotron
photons. Such a mechanism is enormously faster than Coulomb interactions,
overcoming any problem related to fast variability. 

\section{Modeling the Inner Region}

Once argued that thermal IC is the main radiation mechanism at high energies in
Seyfert galaxies, the next question arising is: what is the relation of the
X--ray source to the cold matter responsible for the UV emission? Here I
briefly review four models proposed in the last years for the innermost part of
AGNs. The cartoon version of these models is sketched in Fig.1. 

\begin{figure}
%\vskip -9 true cm
\epsfysize=9cm % fix the y-dimension and scales x-dim. to y-dim.
% Feel free to do the choice you prefer but do not exceed the x-dimension
% of the text lines
\hspace{2.5cm}
\epsfbox{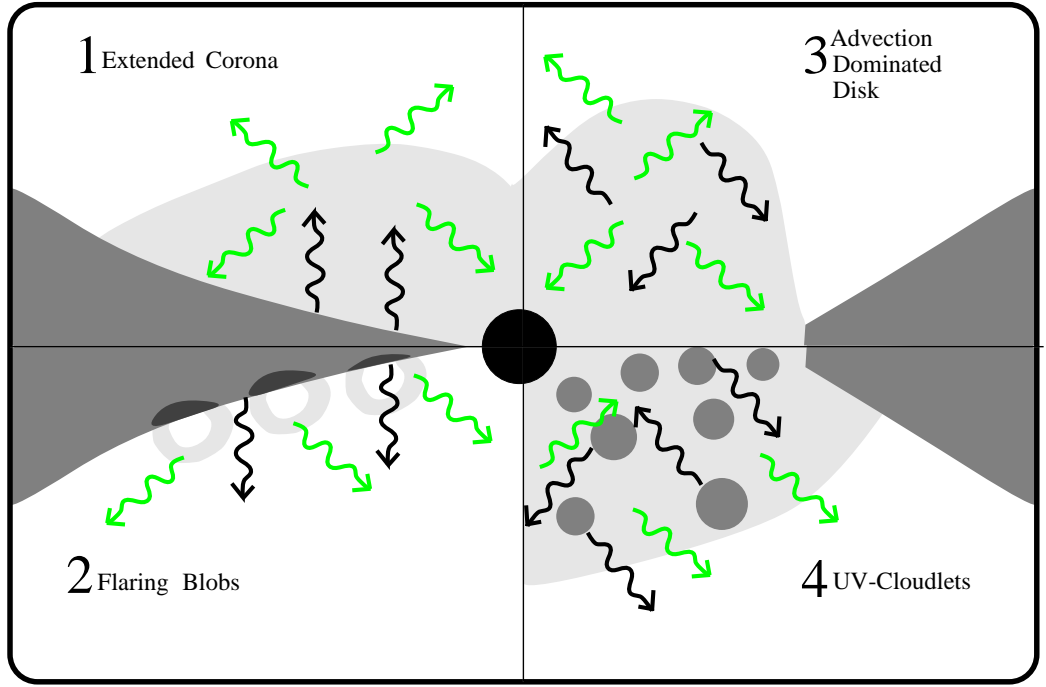} %centering:act on the hspace argument
%\vskip -12. true cm
\caption[h]{The innermost region of radio quiet AGNs according to four current
models for the UV-to-X ray emission. The relative scale of different components
is totally arbitrary. Dark arrows indicate seed photons for Comptonization (IR
to EUV depending on models), while lighter arrows indicate IC emission
(X--rays). See text for model details.} 
\end{figure}

\subsection{Extended Coronae}

In Haardt \& Maraschi (1991, 1993) we developed a  model based on thermal
Comptonization. The hot electrons are thought to be located immediately above
the cold reflecting matter so that the soft blackbody photons emitted by the
cold layer provide the main source of cooling for the hot electrons. At the
same time the hard photons produced by Compton scattering in the hot corona are
an important source of heating for the underlying cold layer which reprocesses
them into soft photons. This interplay between the two phases is the basic
feature of the model. 

We found that the observed spectral features could be explained if the
hard-to-soft luminosity ratio {\it within the corona} is $\sim 2$. In fact such
a value implies a Compton parameter close to 1, that in turn leads to a power
law index close to 1. This value is naturally obtained if one assumes an
extended corona where the whole gravitational power is released, the cold disk
working as a mere reprocessor. For any value of the electron density of the
corona, the temperature adjusts to maintain the same luminosity ratio and so
the same spectral index in the emitted spectrum. 

The optical depth of the corona is in turn fixed by the compactness parameter,
and the resulting equilibrium temperature is tantalizingly close to the derived
value from OSSE data (e.g. Madejski et al. 1995). 

\subsection{Flares above the Disk}

The crucial assumption in the extended corona model above is that all the
available gravitational power is dissipated in the hot corona, the cold
accretion disk acting as a mere reprocessing and reflecting layer. Though this
idea gained recently theoretical support (see Balbus et al. 1996), we may wish
to loose this tight constrain. This can be done considering small scale flaring
regions above the surface of the accretion disk rather than a smooth extended
corona. 

The basic idea is that the disk magnetic field can drain part of the accretion
power outside the accreting flow. In the picture described by Galeev et al.
(1979), the disk differential rotation causes the azimuthal magnetic field to
grow exponentially, if a feedback mechanism operates in order to link the
radial component of magnetic field {\bf B} to the azimuthal one, $B_{\phi}$.
The authors showed that disk convection could be a reliable way to do the job.
In general, the growth of $B_{\phi}$ stops when non linear effects become
important, or because of reconnection of the field lines. The latter mechanism
is unlikely to be effective within the disk, and it can be shown that the
magnetic field can reach the pressure equilibrium with the surrounding gas,
with a subsequent buoyancy of the magnetic flux tubes. In the much more
rarefied coronal ambient magnetic reconnection occurs rapidly because of the
increased Alfv\`en velocity, so that the stored magnetic energy can be
transferred to particles, that in turn radiate. 

The emitted spectrum from the active blobs depend on the dominant energy loss
mechanism effective in cooling the hot particles. One should compare the energy
losses due to bremsstrahlung, cyclotron, synchrotron, inverse Compton radiation
and thermal conduction to the lower cold gas, finding in which regimes IC
dominates. In Haardt et al. (1994) we showed that the conditions of single
blobs are the same as the extended corona (and hence match observations)
provided that the time needed to transfer the accumulated magnetic energy to
the particles is much shorter than the dynamo time scale. 

The cooling of the blob plasma is provided by the reprocessed soft photon flux
below the blobs. The temperature of the reprocessed radiation can be much
higher than that due to the inner disk dissipation, so that the temperature of
the thermal radiation below the blobs will be in general higher than the
temperature of the disk emission at the same radius, producing a hotter thermal
component superimposed to the multicolor disk emission. In this frame the seed
photons for Comptonization in the blobs should be related to the observed
EUV--Soft X--ray excesses rather than to the UV bump. Long term variability can
be ascribed to variations of the accretion rate, while short term variability
can be due to stochastic noise in the number of flaring blobs, that do not need
to be related to the accretion rate. 

\subsection{Advection Dominated Accretion Disks}

The new class of so--called Advection Dominated (accretion) Disks (ADDs) has
been proposed to explain the X--ray emission from underluminous AGNs (e.g. Chen
et al. 1995, and references therein), though with a somewhat extreme choice of
parameters the model could be relevant for "normal" Seyfert galaxies as well
(Narayan \& Yi 1995). 

In ADDs the basic underlying idea is that Coulomb collisions are not fast
enough to transfer energy from ions (supposed to be directly energized by the
accretion process) to electrons. Ions require a catalyzer (the electrons) to
produce photons and hence to cool down. If the interaction between the two
species is slow, ions "advect in" part of the energy gained as they fall into
the hole. This happens if the density of the accreting material is low, i.e. at
low accretion rates. Note that on the other extreme, i.e. at very high
accretion rates, the density can be so high that photons are trapped into the
flow. Also in this case, but for completely different reasons, advection sets
in. The gas is however cold, and not interesting in the context of X--ray
emission. 

On the low accretion rate branch, the whole process can be best understood in
terms of timescales. The three competing timescales are the viscous timescale
$t_{visc}$, the Coulomb timescale $t_{p-e}$ and the cooling timescale for
electrons $t_{cool}$. Viscous processes provide the heating term for ions,
while Coulomb collisions are at the same time the heating term for electrons
and the cooling term for ions. While ions are heated at a constant rate (given
the input disk parameters), electron heating and ion and electron cooling rates
are proportional to the energy content of the gas. Under typical conditions, in
the inner part of the flow we have 
\begin{equation}
t_{cool} \ll t_{visc} \ll t_{p-e}
\end{equation}
The above relations immediately show that a) energy is advected in by protons
(or more generally ions), and b) the plasma is at two temperatures, with
$T_e<T_p<(m_p/m_e)T_e$. Once heated, electrons cool down very rapidly. 

The main result of the conditions described above is that in ADDs the accreting
flow is itself extremely hot, with temperatures of few hundreds keV, and hence
it is directly responsible for the X--ray emission. We may say that the disk
plays the role of the extended corona, with the main difference that in ADDs
there is no cold matter whatsoever (at least in the innermost regions). The hot
disk cools through IC losses. The soft photons to be Comptonized are provided
by internal S and FF emission, rather than by an external soft photon input as
in the extended corona model. S emission, though mostly self-absorbed, is
probably more intense than FF. From the point of view of radiation mechanisms,
we may say that ADDs are then SSC thermal models. We finally note that in ADDs
there are no apparent reasons leading the Compton parameter to be constant
close to 1, so this class of models may have difficulties in explaining the
small dispersion of the [2-20] keV spectral index distribution observed in
Seyfert I galaxies (Nandra \& Pounds 1994). Also the reflection component and
the iron line at 6.4 keV are not well explained. 

\subsection{EUV Cloudlets}

The three models described so far assume that the accreting matter is in an
accretion disk fashion (though in ADDs it turns out to be geometrically thick).
In the first two examples the X--rays are produced in a hot atmosphere
(extended or localized) located above the cold gas, while in ADDs the disk
itself is so hot to become the X--ray source. 

A radically different view was originally proposed by Celotti et al. (1992),
and studied in details by Kuncic et al. (1996). The hot X--ray atmosphere is
supposed to be extended, but unlike ADDs, it is filled with small cold clouds
responsible for the EUV thermal emission, possibly providing soft photons for
IC emission in the hot phase. 

In the cloudlet model, the size of the clouds, which are probably confined by
magnetic fields, can be as small as few meters. These small cloudlets are
supposed to be Thomson thin, heated by FF absorption, and cooled via
collisionally--excited line radiation and FF emission, which process dominates
depending on temperature and density. The cloudlets can account for only part
of the observed UV bump, as they emit the bulk of radiation in EUV lines and
continuum. The authors argue that the spectral energy distribution of radio
quiet AGNs may indeed reach a maximum in the EUV band, rather than in the
optical--UV (for a criticism to this point see Haardt \& Madau 1996). 

In the model version proposed by Collin--Souffrin et al. (1996), the clouds are
supposed to be Thomson thick. In this case the main source of radiation is
free--bound emission, and, depending on the cloudlet filling factor, the
emission can potentially account for the whole UV bump. 

Both the groups focused their studies on the thermal emission from the clouds,
rather than on the X--rays produced in the hot medium. On the framework of
combined UV-to-X ray emission, the interplay between the hot gas and the cold
cloudlets needs to be studied in detail. 

\section{Conclusions}

In the previous sections I have schematically reviewed some aspects of the
problems relative to the X--ray emission from Seyfert galaxies. After giving
evidences of the thermal nature of the IC emission from this class of sources,
I described four current models for the inner region. 

With the present state of the models, and on the basis of currently available
observations, it is difficult to favour one model with respect to the others.
In principle discriminating observations could come from timing properties of
X--ray light curves, comparing properties as time--lags, power spectra and
coherence in different X--ray bands (see e.g. Vaughan \& Nowak 1996). As a
matter of fact most of the information at high energies is expected to come
from observations of Galactic black hole candidates (GBHCs). In this
perspective, models tailored to fit the larger size of AGNs will possibly arise
from interpretation of Galactic data. 

\acknowledgements
I am happy to thank the LOC and all the people involved for the perfect
organization and the great time we all had in Riccione (despite of the flood).

\end{document}